\documentclass[aps,reprint,twocolumn,pra,amsmath,amssymb]{revtex4-1}

\usepackage{natbib}
\usepackage[latin9]{inputenc}
\usepackage{amsmath}
\usepackage{amssymb}
\usepackage{graphicx}
\usepackage{esint}
\usepackage{braket}
\usepackage{multirow}
\usepackage{dcolumn}
\usepackage{color}

\makeatletter
\makeatletter

\begin{document}
\title{Frequency stabilization of a 650 nm laser to I$_{2}$ spectrum for trapped $^{138}$Ba$^{+}$ ions}

\author{Tian Xie$^{1,2}$, Naijun Jin$^{1,2}$, Ye Wang$^{1}$, Junhua Zhang$^{1,3}$, \\ Mark Um$^{1}$, Pengfei Wang$^{1}$, \& Kihwan Kim$^{1}$}

\affiliation{$^{1}$Center for Quantum Information, Institute for Interdisciplinary Information Sciences, Tsinghua University, Beijing 100084, P. R. China  \\ $^{2}$Department of Physics, Tsinghua University, Beijing 100084, P. R. China \\ $^3$ Shenzhen Institute for Quantum Science and Engineering, and Department of Physics, Southern University of Science and Technology, Shenzhen, P. R. China}

\begin{abstract}
The optical manipulation of Ba$^{+}$ ions is mainly performed by a 493 nm laser for the S$_{1/2}$-P$_{1/2}$ transition and a 650 nm laser for the P$_{1/2}$-D$_{3/2}$ transition. Since the branching ratio between the 493 nm and 650 nm transitions of a single Ba$^{+}$ ion is comparable, stabilization systems of both lasers are equally important for Doppler cooling, sub-Doppler cooling, optical pumping and state detection. The stabilization system of a 493 nm laser to an absolute Te$_2$ reference has been well established. However, the stabilization of a 650 nm laser has not been presented before. Here we report twenty spectral lines of I$_{2}$ in the range of 0.9 GHz above the resonance of the P$_{1/2}$-D$_{3/2}$ transition. We stabilize the 650 nm laser through the optical cavity to the lowest one among these lines, which is about 350 MHz apart, as the absolute frequency reference. Furthermore, we measure the frequency differences between these iodine lines and the Ba$^+$ resonance through fluorescence excitation spectrum with well-resolved dark states, which is in agreement with the theoretical expectation. The presented stabilization scheme enables us to perform precise experiments with Ba$^{+}$ ions.
\end{abstract}

\maketitle

\begin{figure*}[htbp]
	\centering
	\includegraphics[width=0.8\textwidth]{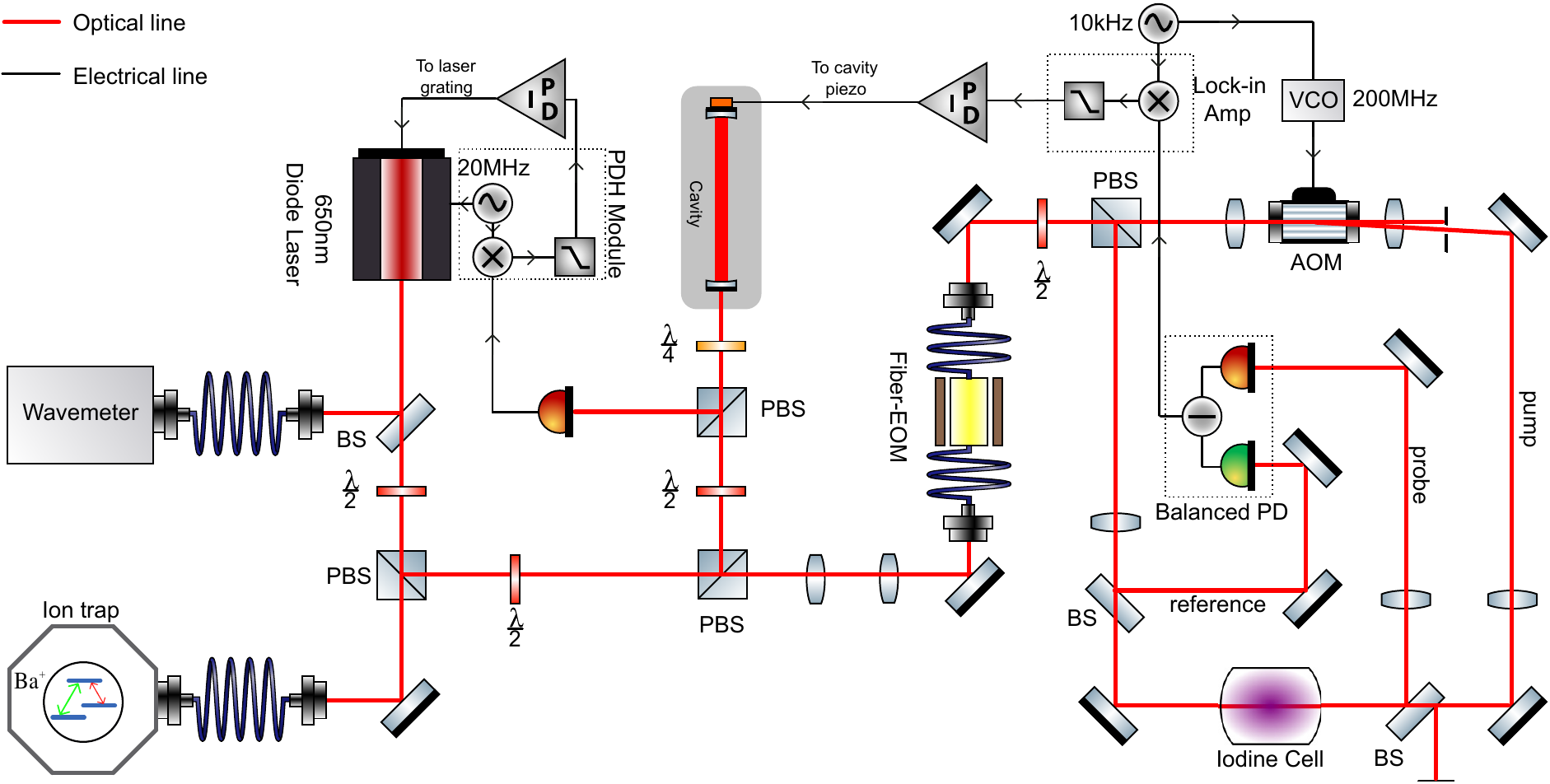}
	\caption{Frequency stabilization system of the 650 nm laser to I$_2$ reference through an optical cavity. The red thick lines show the optical path and the black thin lines with arrows indicate electrical connections. Here 493 nm laser system is not included.}
	\label{setup}
\end{figure*}

\section{Introduction}
Recently, ion trap systems have been greatly developed for quantum computation \cite{cirac1995quantum,haffner2008quantum,monroe2013scaling}, quantum simulation \cite{schneider2012experimental,blatt2012quantum} and fundamental tests of quantum mechanics \cite{kirchmair2009state,chou2010optical,zhang2013state}, where accurate and precise control is required. Since the states of ions are mainly manipulated by optical methods, it is essential to stabilize lasers, particularly in the frequency domain. The performance of a laser system is evaluated by its linewidth and long-term drift of the frequency. To narrow down the laser linewidth, a high-finesse cavity is typically used, for example in the Pound-Drever-Hall (PDH) method \cite{drever1983laser}, but the long-term drift cannot be overcome in this way. 

As a solution, atomic or molecular cells are usually used as the absolute reference for laser stabilization. In particular, hyperfine-resolved optical transitions of iodine (I$_2$) have been thoroughly studied\cite{velchev1998dense,xu2000dense}, providing absolute references from the dissociation limit at 499.5 nm to the near-infrared region including the stabilization of 532, 605, 612, 633, 740 and 830 nm lasers \cite{hall1999stabilization,bertinetto1987helium,cerez1979helium,lazar2000tunable,ludvigsen1992frequency,olmschenk2007manipulation}. Typically, Doppler-free spectroscopy, such as the saturated absorption spectroscopy (SAS) \cite{preston1996doppler} or modulation transfer spectroscopy (MTS) \cite{mccarron2008modulation}, is widely used in laser stabilization systems, where the width of the error signal in the servo-loop can be compressed to tens of MHz level.  

In the field of quantum computation \cite{slodivcka2013atom,wright2015scalable}, quantum communication \cite{auchter2014ion} and precision measurement of parity non-conservation \cite{fortson1993possibility} with trapped ions,  the Ba$^{+}$ ion has been an attractive choice. A single Ba$^{+}$ ion has low-lying and long-lived D states \cite{sherman2005progress} and a lambda structure of cooling cycle \cite{shu2009trapped}, which are all in the visible wavelength range. 
A 493 nm laser drives the S$_{1/2}$-P$_{1/2}$ transition, which performs as the main circulation of Doppler cooling and since the branching ratios of the P$_{1/2}$-S$_{1/2}$ and P$_{1/2}$-D$_{3/2}$ transitions are comparable, as 0.73 and 0.27 \cite{munshi2015precision} respectively, a 650 nm laser bridging the D$_{3/2}$-P$_{1/2}$ transition is in need to pump the population in D$_{3/2}$ state back to the cooling cycle. The 493 nm laser stabilization system with tellurium (Te$_2$) as the absolute reference has been well-developed \cite{raab1998diode}. However, the 650 nm laser has been used without the absolute stabilization for the past years, which could be due to the fact that a modest Doppler-cooling efficiency can be achieved by using a far red-detuned repumping laser. Still the absolute stabilization of the 650 nm repumping laser should be equally important as that of the 493 nm laser, especially when we try to achieve the optimal cooling efficiency and improve the efficiency of optical pumping and quantum-state detection \cite{matthew2009}, which are sensitive to the choice of the detuning of the 650 nm laser. Furthermore, sub-Doppler cooling by EIT or Sisyphus mechanism would critically require the stability of the 650 nm laser at an exact frequency \cite{lechner2016electromagnetically,ejtemaee20173d}. The stabilization system of the 650 nm laser, however, has not been completely discussed \cite{matthew2009,huber2014optical}.

In this work, we report 20 spectral lines of I$_{2}$ close to the P$_{1/2}$-D$_{3/2}$ transition of the $^{138}$Ba$^{+}$ ion at 649.87 nm from the SAS. Furthermore, by employing one of these lines as the absolute reference, we accomplish the absolute stabilization system of the 650 nm laser with the linewidth of $\sim$400 kHz. We also measure the spectrum of a single $^{138}$Ba$^{+}$ ion by scanning the frequency of the 650 nm laser with a fiber-coupled EOM while fixing the frequency of the 493 nm laser and compare the results with the numerical simulation of the Liouville equation \cite{oberst1999resonance}. We conclude that the frequency of the reference line is 350 MHz higher than the resonance of the P$_{1/2}$-D$_{3/2}$ transition of the $^{138}$Ba$^{+}$ ion.

\section{I$_{2}$ spectroscopy}
\begin{figure}[htbp]
	\centering
	\includegraphics[width=0.48\textwidth]{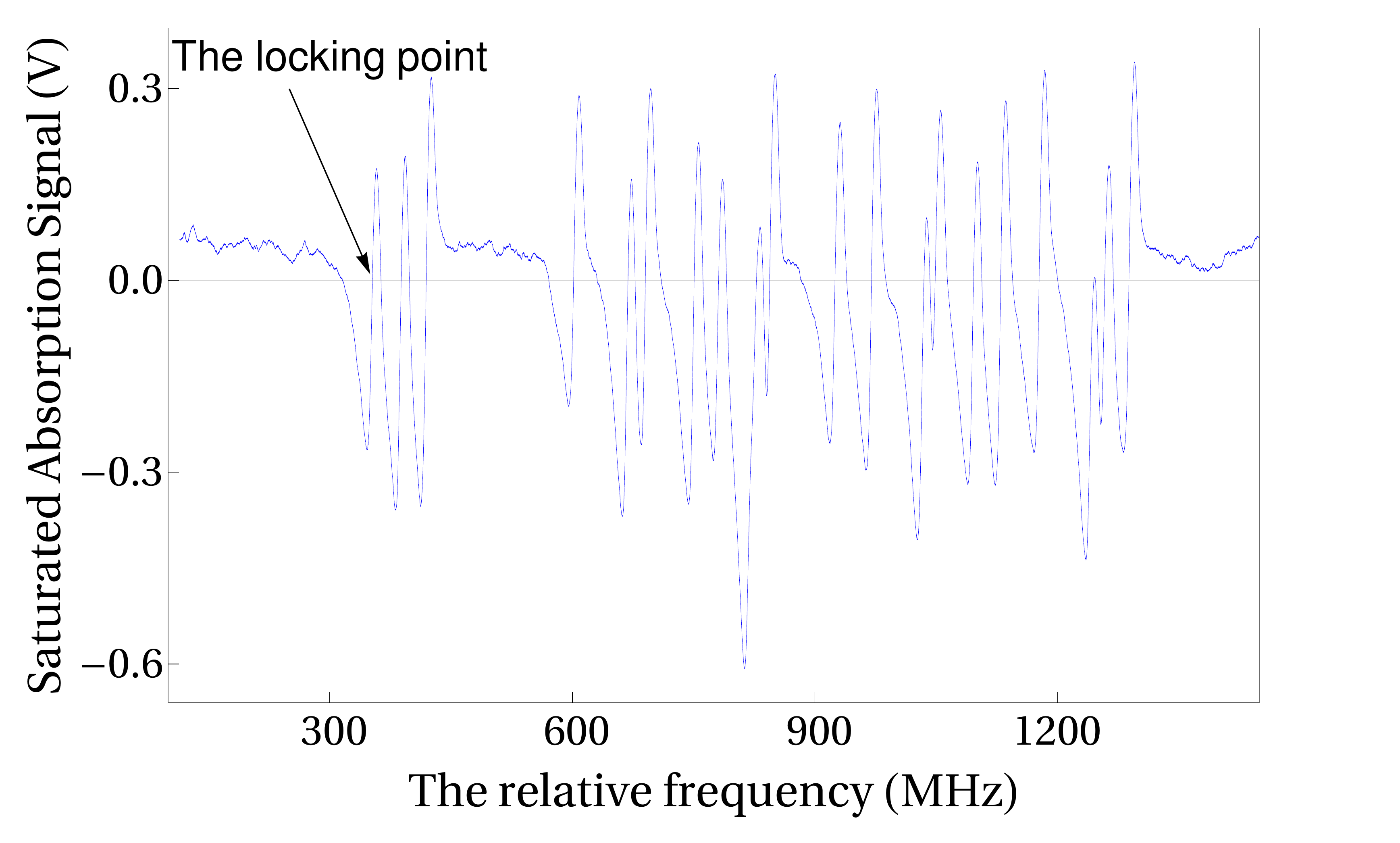}
	\caption{The derivative of the saturated absorption spectroscopy (SAS) of I$_{2}$ near the P$_{1/2}$-D$_{3/2}$ transition of $^{138}$Ba$^{+}$. Here the frequency axis shows the detuning between the absorption lines of I$_{2}$ and the resonance of the transition, which is determined by later experiments. The arrow indicates the locking point of the 650 nm laser.}
	\label{SAS}
\end{figure}

In this section, we introduce the absolute stabilization system of the 650 nm laser and the SAS of I$_{2}$ as shown in Fig. \ref{setup} and Fig. \ref{SAS}. The whole system can be divided into two parts. We first narrow the linewidth of the diode laser Toptica DL100 pro by an optical cavity. To suppress its length drift mainly caused by temperature fluctuation, we then stabilize the optical cavity to the absolute reference of I$_{2}$ by means of SAS. The output of the laser with 18.4 mW power is split into four beams by one beam splitter(BS) and two polarizing beam splitters(PBS). Among these beams, about 0.3 mW is used for the wavemeter from HighFinesse, about 0.9 mW for the ion trap experiment, about 3.5 mW for the optical cavity setup and the rest with about 10 mW is coupled into a fiber-coupled EOM for the implementation of the SAS setup.

Due to the low efficiency of the fiber-coupled EOM, only about 2 mW is available for the optical part of the SAS setup. After passing through a half wave plate and a PBS, the beam is split into two, and by adjusting the half wave plate can we change their relative powers. The transmission beam of the PBS, the pump beam, is sent through an AOM and its +1st order output(430 $\mu$W) is selected. The reflection beam (250 $\mu$W) goes through a convex lens and is then split into a reference beam (70 $\mu$W) and a probe beam (160 $\mu$W). The reference beam goes into one port of the balanced photodiode (balanced PD). The probe beam and the modulated pump beam counterpropagate through an I$_{2}$ cell, which diminishes the effect of Doppler broadening by narrowing the velocity region of I$_2$ that can simultaneously interact with both beams. After that, the probe beam goes into the other port of the balanced PD.

To implement the phase-sensitive detection of our dispersive signal, we incorporate an AOM driven by a frequency-modulated signal from 190 MHz to 210 MHz with a scanning frequency of 10 kHz from a voltage controlled oscillator (VCO). This modulation signal of 10 kHz also performs as the reference of a lock-in amplifier which mixes the PD signal with the reference. We use a servo PID board to process the generated signal and stabilize cavity length by controlling the piezo voltage. 


Fig. \ref{SAS} shows the dispersive signal by scanning the frequency of the repumping laser in a range of 1.2 GHz around 649.8683 nm, which serves as a proper zero-crossing dispersive error signal for servo system. The frequency shifting due to the offset frequency of the modulation signal provided by the VCO has been subtracted and the absolute frequency has been calibrated with the Ba$^+$ ion spectroscopy that will be discussed later. We choose the first line as the absolute reference to perform the frequency stabilization for convenience. 

\section{$^{138}\text{Ba}^{+}$ spectroscopy}
\begin{figure}[htbp]
	\centering
	\includegraphics[width=0.48\textwidth]{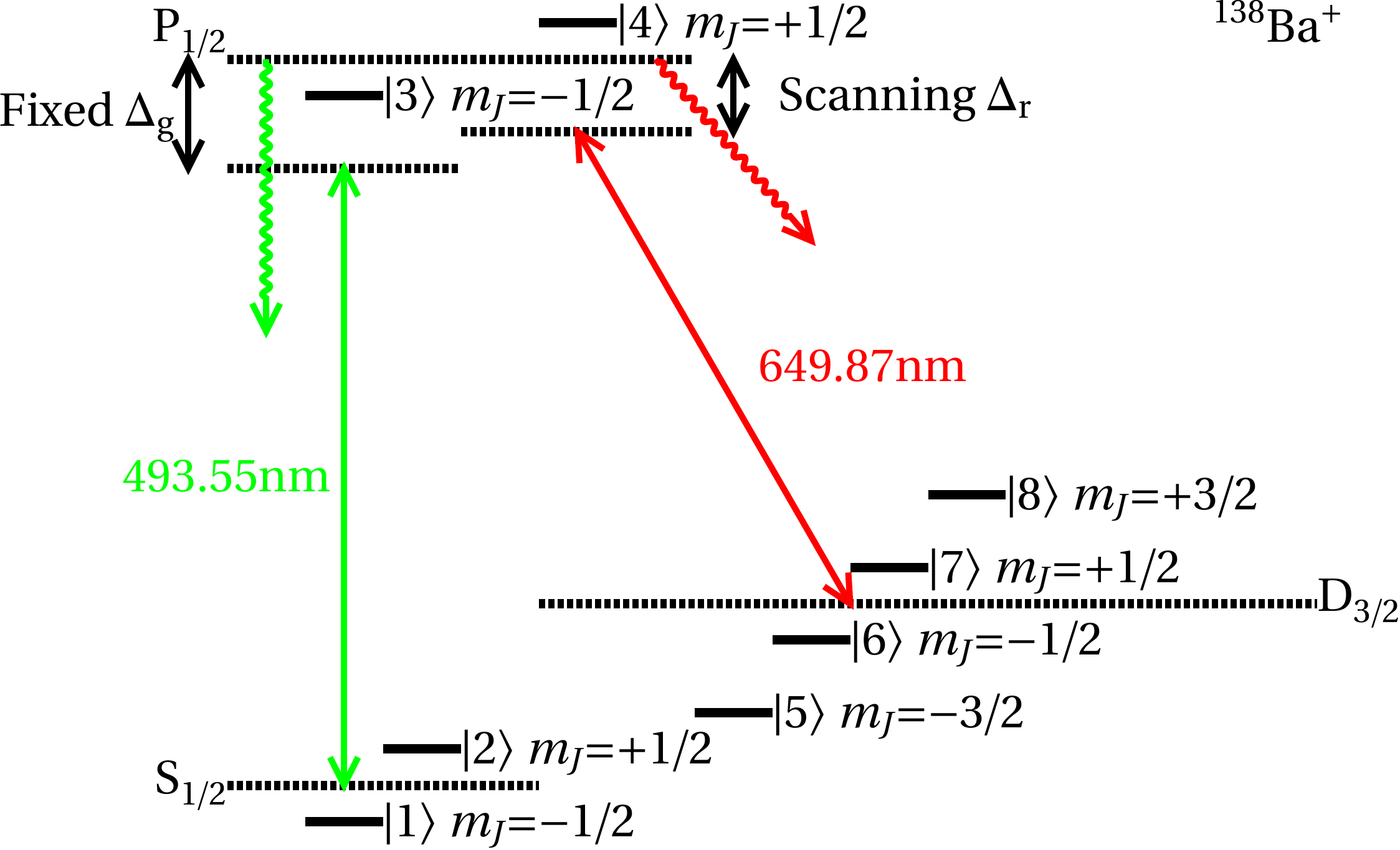}
	\caption{The Energy level of the $^{138}$Ba$^{+}$ ion. In the presence of the magnetic field, the 3-level system splits into an 8-level system. $\Delta_{\rm{g}}$ ($\Delta_{\rm{r}}$) is the 493 (650) nm laser detuning to the resonance of the $^{138}$Ba$^{+}$ ion transition in the absence of Zeeman splitting. We label each state as $\ket{i}$ for the convenience of calculation. 
	\label{structure}
	}
\end{figure}

In order to determine the frequency differences between 20 lines and the resonance of the P$_{1/2}$-D$_{3/2}$ transition of the $^{138}$Ba$^{+}$ ion, we simulate the fluorescence excitation spectrum and compare it with the experimental results. We consider the situation that the detuning of the 493 nm laser is fixed and the detuning of the 650 nm is scanned. Fig. \ref{structure} shows the lowest three energy levels of a single $^{138}$Ba$^{+}$ ion with its Zeeman sublevels. As mentioned in the introduction, the S$_{1/2}$-P$_{1/2}$ transition is the main cooling circulation. However, after excited to the P$_{1/2}$ state, the single ion have a probability of decaying to the D$_{3/2}$ state with a long lifetime, which terminates the cooling cycle. Therefore, we use the 650 nm laser to repump the ion to the P$_{1/2}$ state. The repumping efficiency which is related to the frequency of the 650 nm laser essentially influences the cooling efficiency.  After the loading process, a single $^{138}$Ba$^{+}$ ion in the ground state S$_{1/2}$ is excited to the P$_{1/2}$ state by a 493 nm laser locked to a Te$_{2}$ reference. When scanning the frequency of the 650 nm laser, we measure the intensity of 493 nm fluorescence which can be interpreted as the repumping efficiency of the 650 nm laser as discussed above. In the following sections, we show the theory of fluorescence excitation spectrum and our experimental results compared with the numerical simulation.

\begin{figure}[htbp]
	\centering
	\includegraphics[width=0.3\textwidth]{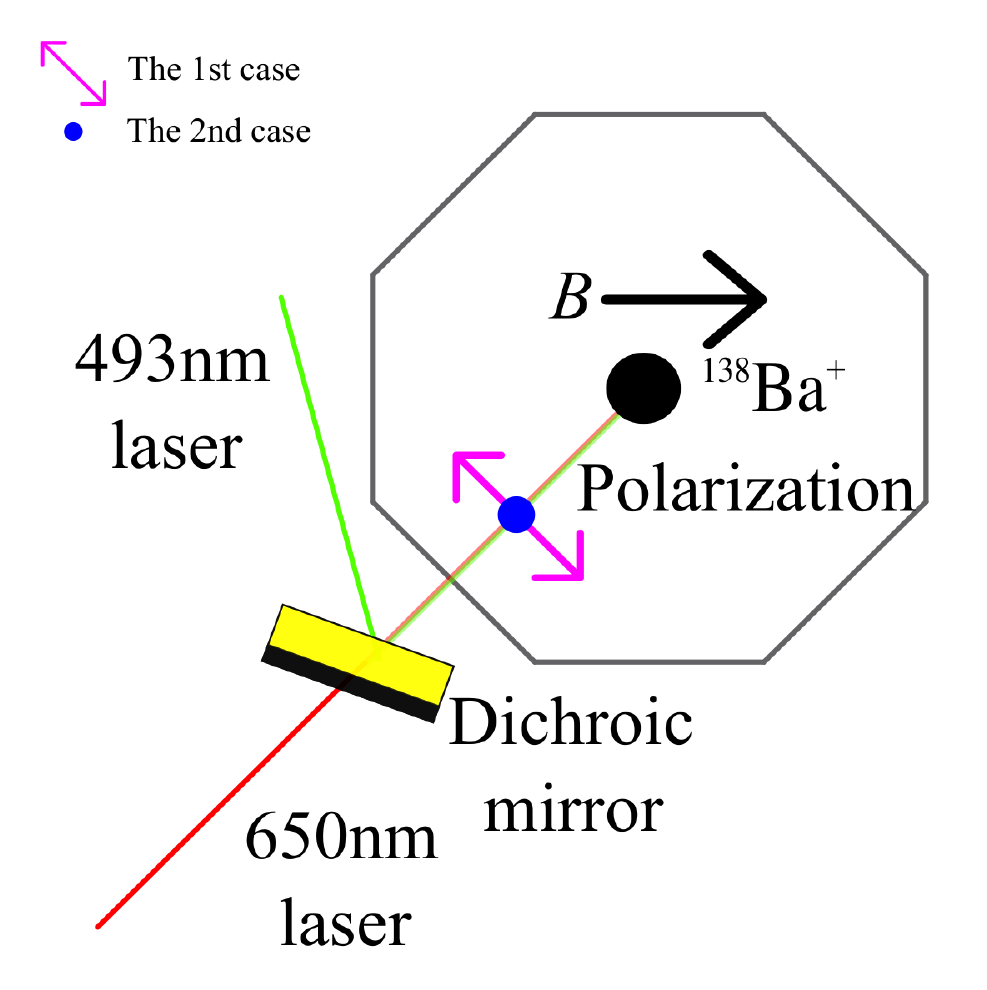}
	\caption{Directions of the lasers and the magnetic field. The first case--the polarizations of both lasers form an angle of $45^{\circ}$ to the magnetic field.The second case--the polarizations are perpendicular to the magnetic field.}
	\label{polar}
\end{figure}

\section{Theory of the fluorescence excitatiton spectrum}

In the presence of 5.8 Gauss magnetic field, the system should be described as an 8-level system \cite{oberst1999resonance}.
By denoting each state shown in Fig. \ref{structure} as $\ket{1}$, $\ket{2}$,...,$\ket{8}$, the atomic Hamiltonian can be expressed as
\begin{equation}
\begin{aligned}
\hat{\mathcal{H}}_{\text{atom}}=\sum_{\rm{i}=1}^{8}\hbar\omega_{\rm{i}}\ket{\rm{i}}\bra{\rm{i}},
\end{aligned}
\end{equation}

As for the interaction term, we treat two lasers as classical electromagnetic waves $\vec{E}_{\rm{g}}\sin(\omega_{\rm{g}}t)$ and $\vec{E}_{\rm{r}}\sin(\omega_{\rm{r}}t)$ because of their high intensity and assume that the light field only interacts with the electric dipole moment of the ion. By applying the rotating-wave approximation, the interaction Hamiltonian remains in the form
\begin{equation}
\begin{aligned}
\hat{\mathcal{H}}_{\text{int}}=&-\frac{i}{2}[\vec{E}_{\rm{g}}\cdot(\vec{D}^{*}_{13}\ket{3}\bra{1}+ \vec{D}^{*}_{14}\ket{4}\bra{1}+\cdot\cdot\cdot)e^{-i\omega_{\rm{g}}t}\\
&+\vec{E}_{\rm{r}}\cdot(\vec{D}_{35}\ket{3}\bra{5}+ \vec{D}_{36}\ket{3}\bra{6}+\cdot\cdot\cdot)e^{-i\omega_{\rm{r}}t}]+h.c..
\end{aligned}
\end{equation}

However, spontaneous decay exists in the system due to the interaction of the ion with the vacuum modes of the light field. A density matrix method should be employed to describe the whole system.
The dynamic of this system is governed by Liouville equation with a decay term \cite{scully1999quantum}
\begin{equation}
\begin{aligned}
\frac{\mathrm{d} \hat{\rho}}{\mathrm{d} t}&=-\frac{i}{\hbar}[\hat{\mathcal{H}},\hat{\rho}]+\mathcal{L}_{\text{damp}}(\hat{\rho})\\
\mathcal{L}_{\text{damp}}(\hat{\rho})&=-\frac{1}{2}\sum_{\rm{m}}[\hat{C}^{\dagger}_{\rm{m}}\hat{C}_{\rm{m}}\hat{\rho}+\hat{\rho}\hat{C}^{\dagger}_{\rm{m}}\hat{C}_{\rm{m}}-2\hat{C}_{\rm{m}}\hat{\rho}\hat{C}^{\dagger}_{\rm{m}}].
\end{aligned}
\end{equation}
The general form of $\hat{C}_{\rm{m}}=\sqrt{\Gamma_{\rm{ij}}}\ket{\rm{j}}\bra{\rm{i}}$ could summarize several different dissipative processes from $\ket{\rm{i}}$ to $\ket{\rm{j}}$ where $\Gamma_{\rm{ij}}$ is the decay parameter, including spontaneous decay from $P_{1/2}$ to $S_{1/2}$ or $D_{3/2}$ and the decoherence due to the finite linewidth of the driving light fields.
In the experiment, we detect the fluorescence photon number every 1 ms, in which case the system has already evolved into a steady state, and thus, all the elements in density matrix could be numerically solved by
\begin{equation}
\begin{aligned}
\frac{\mathrm{d} \hat{\rho }}{\mathrm{d} t}=0\Rightarrow\frac{i}{\hbar}[\hat{\mathcal{H}}&,\hat{\rho}]=\mathcal{L}_{\text{damp}}(\hat{\rho})\\
tr(\hat{\rho})&=1
\end{aligned}
\end{equation}
The fluorescence is proportional to $\rho_{33}+\rho_{44}$, the population of P$_{1/2}$ level, where the ratio depends on the fluorescence collection efficiency. When the detunings of two lasers are exactly the same, which is the condition for Raman transition between S$_{1/2}$ and D$_{3/2}$ state, the P$_{1/2}$ state won't be populated and hence no fluorescence can be detected. This effect, which is called dark resonance, results in a deep hole in the spectrum of the ion when scanning the frequency of one laser and fixing the other. Under such condition, the system will stay in a superposition state between some specific Zeeman sublevels of S$_{1/2}$ and D$_{3/2}$ state, named dark state, which could be seen clearly by calculating the off-diagonal terms in density matrix. 

\section*{Measurement and results}

\begin{figure}[htbp]
	\centering
	\includegraphics[width=0.45\textwidth]{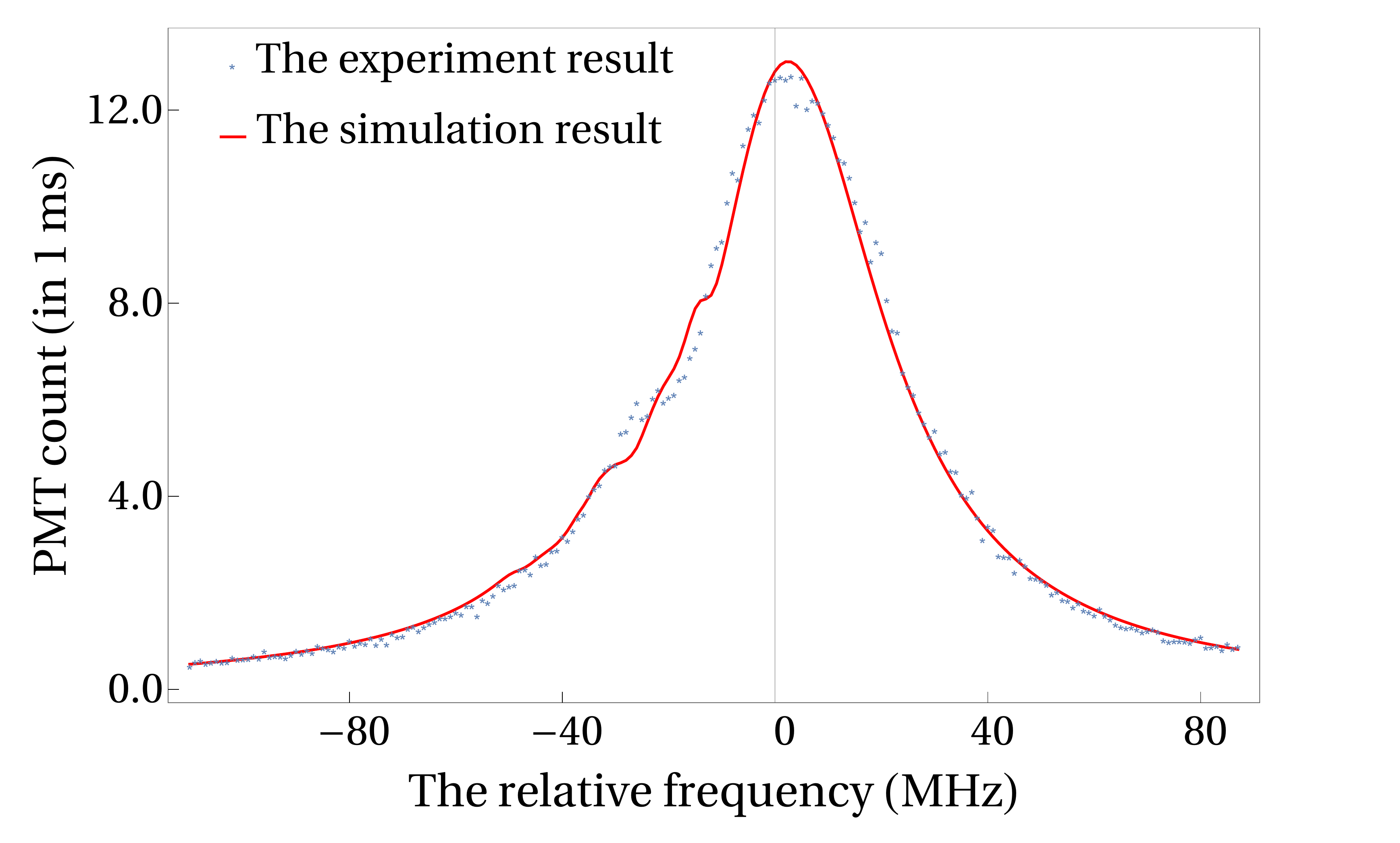}
	\caption{
		The fluorescence excitation spectrum of a single $^{138}$Ba$^{+}$ ion by scanning the frequency of the 650 nm laser. Parameters are $\Delta_{\rm{g}}/2\pi$ = -30 MHz, $B$ = 5.8 G, 493 nm power = 1.5 $\mu$W, 650 nm power = 1.2 $\mu$W, $\alpha$ = $45^{\circ}$, 493 linewidth = 0.3 MHz, 650 linewidth = 0.4 MHz.
	}\label{GaussianLike}
\end{figure}

\begin{figure}[htbp]
	\centering
	\includegraphics[width=0.45\textwidth]{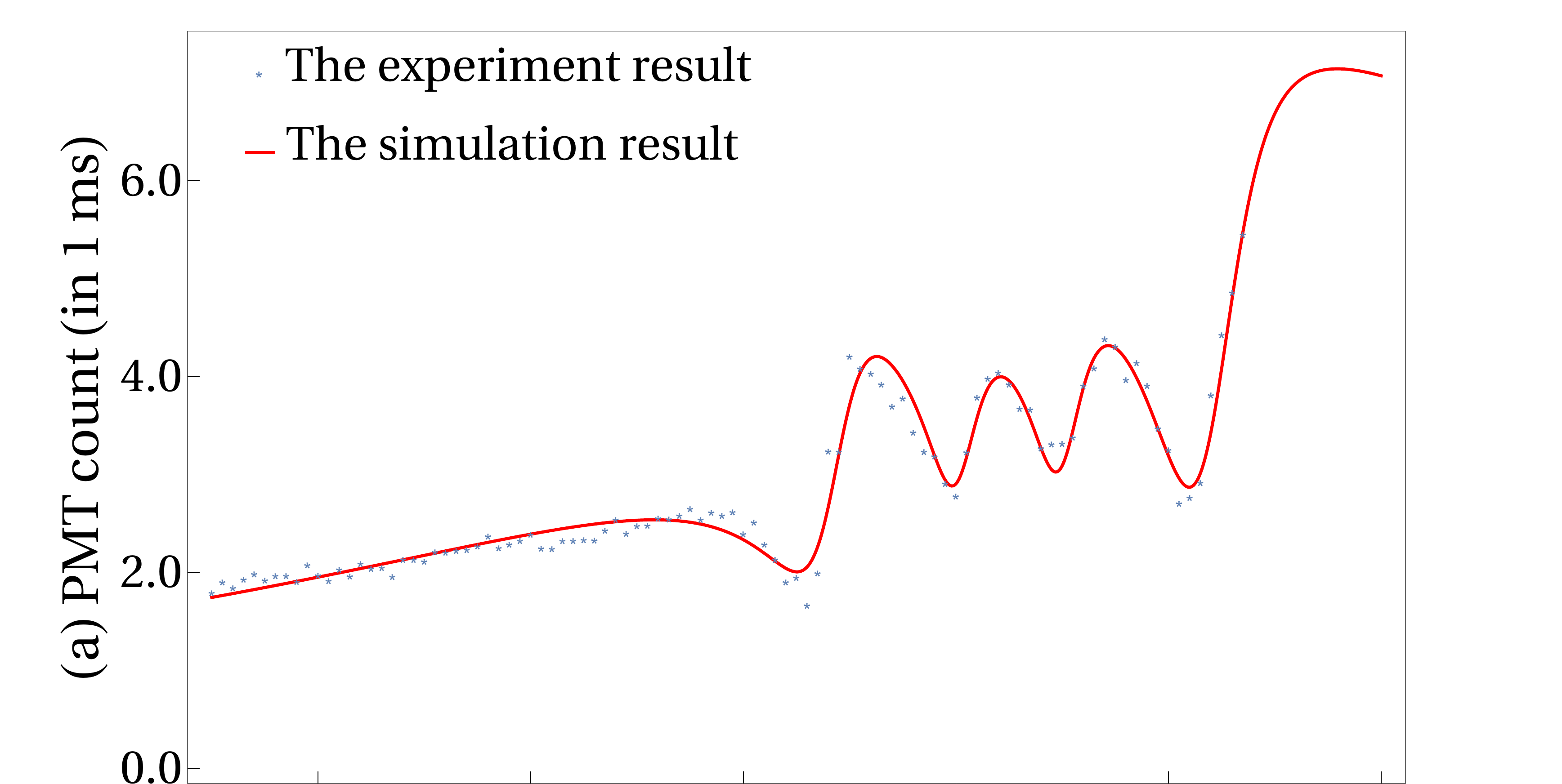}
	\includegraphics[width=0.45\textwidth]{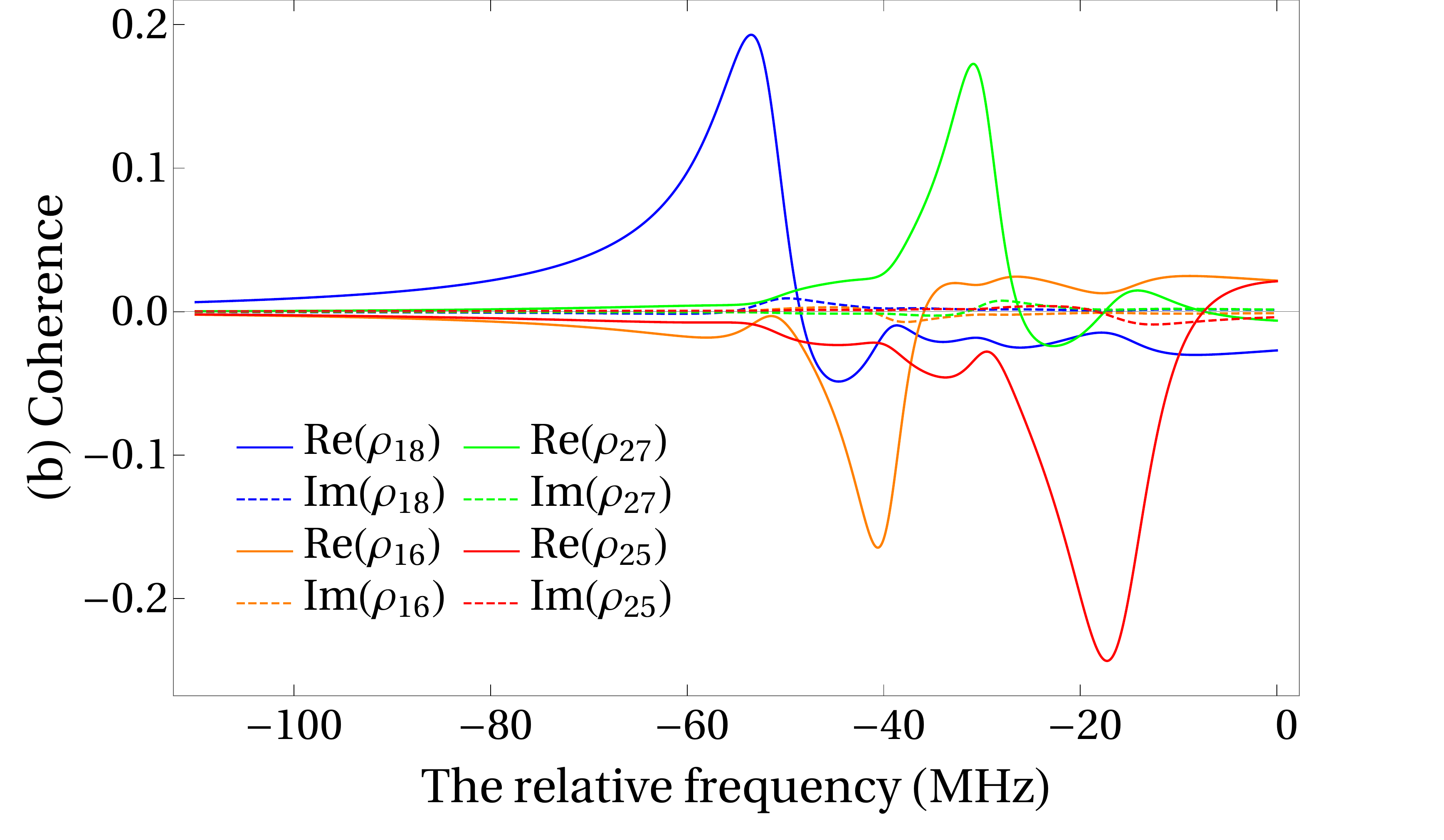}
	\caption{
		The fluorescence excitation spectrum of a single $^{138}$Ba$^{+}$ ion by scanning the frequency of the 650 nm laser. Parameters are $\Delta_{\rm{g}}/2\pi$ = -35 MHz, $B$ = 5.8 G, 493 nm power = 3.7 $\mu$W, 650 nm power = 12 $\mu$W, $\alpha$ = $90^{\circ}$, 493 linewidth = 0.3 MHz, 650 linewidth = 0.4 MHz. When the system is in the dark state, the ion will stay in a superposition state between some specific Zeeman sublevels of S$_{1/2}$ and D$_{3/2}$ state, which could be seen clearly by calculating the off-diagonal terms in density matrix.
	}\label{DarkState}
\end{figure}

In the experiments, we keep the same linear polarization for both laser beams and use $\alpha$ to denote the angle between the polarization and the direction of the magnetic field. In the first case shown in Fig. \ref{polar}, $\alpha$ is $45^{\circ}$ with the 493 nm laser power of 1.5 $\mu$W and 650 nm laser power of 1.2 $\mu$W. We scan the frequency of the 650 nm laser while fixing the frequency of the 493 nm laser and obtain a spectrum close to the Gaussian shape, as shown in Fig. \ref{GaussianLike}. We find the frequency difference between the resonance of the P$_{1/2}$-D$_{3/2}$ transition and the locking point by fitting Fig. \ref{GaussianLike} with the simulation curve. For further tests, we adjust the laser polarization perpendicular to the magnetic field shown as the second case in Fig. \ref{polar} and obtain spectrums which contain characteristic dark states. Because there are only $\sigma_{+}$ and $\sigma_{-}$ interactions in the second case, the dark state will occur when the detuning of the 650 nm laser satisfies
\begin{equation}
\begin{aligned}
\Delta_{\rm{r}} = \Delta_{\rm{g}} \pm3/5\frac{\mu_{\rm{B}}\left | \vec{B} \right |}{h}, \Delta_{\rm{g}} \pm11/5\frac{\mu_{\rm{B}}\left | \vec{B} \right |}{h}. 
\label{dresonance}
\end{aligned}
\end{equation}
And in the case where the power of 493 nm and 650 nm lasers are 3.7 $\mu$W and 12 $\mu$W respectively, the result compared with the simulation is shown in Fig. \ref{DarkState}(a).
Fig. \ref{DarkState}(b) shows the coherence terms, the off-diagonal elements in density matrix between some specific sublevels of S$_{1/2}$ and D$_{3/2}$. In the experiment, we observe the dark resonances occur at -54 MHz, -40 MHz, -31 MHz, and -18 MHz, which are in agreement with the theoretical expectation of Eq. (\ref{dresonance}).

From these experimental results, we find that the resonance frequency of P$_{1/2}$-D$_{3/2}$ transition is 350 MHz lower than the frequency of the first line of I$_{2}$ line in Fig. \ref{SAS}, which is close enough for an AOM to bridge in further experiments. With this conclusion, center frequencies of these twenty lines of I$_2$ are listed in Table \ref{table} in terms of the relative value to the resonance of the P$_{1/2}$-D$_{3/2}$ of a single $^{138}$Ba$^{+}$ ion.
~\\
\begin{table}[htbp]
\centering
\begin{tabular}{cccc}
\hline
Num&Relative Freq(MHz)&Num&Relative Freq(MHz)\\
\cline{1-4}
1&350&11&925\\
\cline{1-4}
2&386&12&969\\
\cline{1-4}
3&418&13&1033\\
\cline{1-4}
4&600&14&1048\\
\cline{1-4}
5&667&15&1093\\
\cline{1-4}
6&690&16&1128\\
\cline{1-4}
7&751&17&1174\\
\cline{1-4}
8&780&18&1237\\
\cline{1-4}
9&826&19&1258\\
\cline{1-4}
10&844&20&1288\\
\hline
\end{tabular}
\caption{The center frequency of each line relative to the resonance of the P$_{1/2}$-D$_{3/2}$ transition. Those results are obtained by scanning the laser frequency in a small range around each resonance and recording the laser frequency from the wavemeter. Each marked center frequency has an uncertainty of 3 MHz due to the fluctuation of the wavemeter.}
\label{table}
\end{table}

\section{Conclusion}

We have implemented the SAS of I$_2$ around 650 nm for the trapped Ba$^+$ ion experiment. We report twenty lines of I$_2$ in the vicinity of the  P$_{1/2}$-D$_{3/2}$ transition of the Ba$^+$ ion and we have accomplished the stabilization of the 650 nm laser by using one of the lines as the absolute reference. We also have performed the spectroscopy of a single $^{138}$Ba$^+$ ion and concluded that the resonance of the P$_{1/2}$-D$_{3/2}$ transition is 350 MHz lower than the lowest absolute reference line among the observed 20 lines of I$_2$. Our development would enable us to perform precise experiments with Ba$^+$ ion including sub-Dopper cooling for quantum computation or precision measurement.

\section*{Acknowledgment}
This work was supported by the National Key Research and Development Program of China under Grants No. 2016YFA0301900 and No. 2016YFA0301901 and the National Natural Science Foundation of China Grants No. 11374178, and No. 11574002.

%

\end{document}